\documentclass[aps,prl,preprint,groupedaddress,  showkeys]{revtex4-1}
\usepackage{color}
\usepackage{amsmath}
 \usepackage[squaren]{SIunits}
 \usepackage{graphicx}

 \newcommand{\pts}[1] {\left(#1\right)}
\begin{document}









\title{Highly efficient parallel grand canonical simulations of interstitial-driven deformation-diffusion processes}

\author{S. Sina Moeini-Ardakani}
\affiliation{Department of Civil and Environmental Engineering,
  Massachusetts Institute of Technology, 77 Massachusetts Avenue, Cambridge, MA 02139,   USA}
\author{S. Mohadeseh Taheri-Mousavi}
\affiliation{Department of Mechanical Engineering and Department of Materials
  Science and Engineering, Massachusetts Institute of Technology, 77 Massachusetts Avenue, Cambridge, MA 02139, USA}
\author{Ju Li}
  \email[Corresponding author: ]{liju@mit.edu}
\affiliation{Department of Nuclear Science and Engineering and
  Department of Materials Science and Engineering, Massachusetts Institute of
  Technology, 77 Massachusetts Avenue, Cambridge, MA 02139, USA}
\date{\today}

\begin{abstract}
Diffusion of interstitial alloying elements like H, O, C, and N in metals 
and their continuous relocation and interactions with their microstructures 
have crucial influences on metals’ properties. However, besides 
limitations in experimental tools in capturing these mechanisms, the inefficiency 
of numerical tools also inhibits modeling efforts. Here, we present an efficient framework to 
perform hybrid grand canonical Monte Carlo and molecular dynamics simulations that allow for parallel insertion/deletion of Monte Carlo moves. A new 
methodology for calculation of the energy difference at trial moves that can 
be applied to many-body potentials as well as pair 
ones is a primary feature of our implementation. We study H 
diffusion in Fe (ferrite phase) and Ni polycrystalline samples to demonstrate the efficiency and 
scalability of the algorithm and its application. The computational cost of using 
our framework for half a million atoms is a 
factor of 250 less than the cost of using existing libraries.
\end{abstract}


\maketitle
\section{Introduction}


H-embrittlement, oxidation, creep, and carbide precipitation
are examples of life-limiting chemo-mechanical mechanisms of metallic
alloys in service. While continuum theories can capture the
deformation mechanism phenomenologically, they have a limited range of
validity when the inherent interactions between
deformation and diffusion are unspecified. On the other hand, simulations
at atomic resolution can reveal the underlying mechanisms and their
correlations, but are severely limited in computational efficiency.

Molecular dynamics (MD) simulations have been widely used for
equilibrium and non-equilibrium processes, while they have limitations
in capturing long time-scale mechanisms such as diffusion. Monte Carlo
(MC) simulations, besides their convenience in implementation, cannot
capture non-equilibrium deformation mechanisms. The combination of
these two techniques seems to be ideal. However, in practice, the
probabilistic nature of the MC scheme and also the evolving number
of degrees of freedom complicate the efficient implementation, i.e.,
parallelization of the hybrid MC-MD framework on multi-core architectures.

A parallelization scheme that can perform 
simultaneous MC moves based on domain 
decomposition was proposed by Sadigh et al. \cite{sadigh_scalable_2012}. Later on, Yamakov \cite{yamakov_parallel_2016}
implemented parallel MD and semi-grand canonical Monte Carlo (SGCMC) simulations 
using this algorithm to model deformation-diffusion of substitutional alloys. Here, we present a newly developed software package 
that, among many features, has new hybrid grand canonical 
Monte Carlo (GCMC) and MD algorithms to model interstitial deformation-diffusion processes. In this package, GCMC, much like MD, is efficiently 
parallelized using domain decomposition. In 
addition, we introduce a more generalized 
concept of the linked list algorithm \cite{allen_computer_2017} 
that can be utilized to greatly improve the performance of 
the GCMC algorithm. Performing isothermal H-charging and discharging in a Ni model material will show that our library has two orders of magnitude less computational cost than LAMMPS \cite{plimpton_fast_1995}. Moreover, being able to conduct simulations at polycrystalline scale reveals the unknown parameters of analytic formulas to extract the concentration-pressure relationship of different microstructural defects in a model material system.

\section{Methodology}
\subsection{Theory}

\subsubsection{Proof of Detailed Balance}

Consider a simulation supercell with a fixed total volume $V$,
temperature $T$, and a chemical potential $\mu$ of an isotope of mass
$m$.  Even though the treatment here is for a monatomic system, it can
be trivially generalized to multiple chemical species (isotopes), with
constant $\{\mu_c, m_c\}$ for $c=1\cdots C$ nuclide species. In the
semi-classical treatment, the grand partition function $\mathcal{Q}$
can be written for $N$ identical particles of position vector ${\bf
  x}^{3N}\equiv {\bf x}_1{\bf x}_2...{\bf x}_N$ as follows
\cite{kardar_statistical_2007}:

\begin{align}
  \label{GrandPartitionFunction}
  \mathcal{Q}\pts{T,\mu,V} \equiv \sum_{N=0}^{\infty}
  \int \frac{d{\bf x}^{3N}}{N! \lambda_{\mathrm{th}}^{3N}}  e^{\beta\pts{N\mu-U\pts{\mathbf{x}^{3N}}}},
\end{align}
where $\beta\equiv \pts{k_{\mathrm{B}} T}^{-1}$,
$\lambda_{\mathrm{th}}\equiv h/\sqrt{2\pi m k_{\mathrm{B}} T}$ is the
thermal de Broglie wavelength, and $U\pts{\mathbf{x}^{3N}}$ denotes
the potential energy of the system. The differential probability of
finding the system at a phase-space volume $\pts{N,\mathbf{x}^{3N}}$
is therefore:
\begin{align}
\label{pdf:micro}
 p\pts{N,\mathbf{x}}d\mathbf{x}^{3N} =\frac{d\mathbf{x}^{3N}}{N!
   \lambda_{\mathrm{th}}^{3N}} e^{\beta\pts{\mathcal{G}+\mu N-
     U\pts{\mathbf{x}^{3N}}}},
\end{align}
where $\mathcal{G}\equiv -k_{\mathrm{B}}T\log\mathcal{Q}$ is the grand
potential. Note that if we consider the particle-index permutation
symmetry in ${\bf x}^{3N}\equiv {\bf x}_1{\bf x}_2...{\bf x}_N$, there
are $N!$ copies of this phase-space volume with exactly the same $U$
and therefore the same probability density. (\ref{pdf:micro}) represents just
one of these copies in a particular differential volume $d{\bf
  x}^{3N}= d{\bf x}_1d{\bf x}_2...d{\bf x}_N$, namely nuclide 1 in
$({\bf x}_1,{\bf x}_1+d{\bf x}_1)$, nuclide 2 in $({\bf x}_2,{\bf
  x}_2+d{\bf x}_2)$, ..., nuclide $N$ in $({\bf x}_N,{\bf x}_N+d{\bf
  x}_N)$.

Metropolis Monte Carlo\cite{metropolis_equation_1953} relies on
transition rates that respects Detailed Balance.  If we perform
particle insertion with some ${\rm
  transition}\pts{N,\mathbf{x}^{3N}\rightarrow N+1,\mathbf{x}^{3N+3}}
d\mathbf{x}^{3N+3}$ rate where the proportionality to destination
volume $d\mathbf{x}^{3N+3}$ is made explicit, leaving the ${\rm
  transition}\pts{N,\mathbf{x}^{3N}\rightarrow N+1,\mathbf{x}^{3N+3}}$
function itself an intensive quantity with finite value, then the
corresponding reciprocal deletion rate must satisfy
\begin{equation}
  \label{DetailedBalanceRequirement}
 \frac{{\rm transition}\pts{N,\mathbf{x}^{3N}\rightarrow
   N+1,\mathbf{x}^{3N+3}} d\mathbf{x}^{3N+3}}{{\rm
     transition}\pts{N+1,\mathbf{x}^{3N+3}\rightarrow N,\mathbf{x}^{3N}
 } d\mathbf{x}^{3N}} \;=\;
 \frac{p\pts{N+1,\mathbf{x}^{3N+3}}d\mathbf{x}^{3N+3}}{p\pts{N,\mathbf{x}^{3N}}d\mathbf{x}^{3N}},
\end{equation}
where the right-hand side (RHS) is the ratio of the resident
probabilities that we desire to approach, and the left-hand side (LHS)
is the ratio of the conditional transition probabilities. It is then
clear that differential phase-space volumes $d\mathbf{x}^{3N+3}$,
$d\mathbf{x}^{3N}$ be cancelled out from both sides, making the exact
values of these infinitesimal quantities immaterial, as they should
be.

In principle, there needs to be no relation between $\mathbf{x}^{3N}$
and $\mathbf{x}^{3N+3}$; in other words, the positions of all atoms
can be changed, even by a lot, in one move. But in the simplest
incarnation that preserves Detailed Balance, we choose to preserve
almost all of the atomic positions except for the atom (or a position)
in question:
\begin{equation}
  \mathbf{x}^{3N+3} \;=\; \mathbf{x}^{3N} {\bf x}_{\rm question}
\end{equation}
in which case 
\begin{equation}
  \label{DifferentialVolume}
 d\mathbf{x}^{3N+3} \;=\; d\mathbf{x}^{3N} d{\bf x}_{\rm question},
\end{equation}
and (\ref{DetailedBalanceRequirement}) is simplified to
\begin{equation}
  \label{DetailedBalanceRequirement2}
 \frac{{\rm transition}\pts{N,\mathbf{x}^{3N}\rightarrow
   N+1,\mathbf{x}^{3N+3}} d{\bf x}_{\rm question} }{{\rm
     transition}\pts{N+1,\mathbf{x}^{3N+3}\rightarrow N,\mathbf{x}^{3N}
 } } \;=\;
 \frac{e^{\beta\pts{\mu -
     \Delta U}} d{\bf x}_{\rm question}}{(N+1)
   \lambda_{\mathrm{th}}^{3}}
\end{equation}
where 
\begin{equation}
 \Delta U \;\equiv\; U\pts{\mathbf{x}^{3N+3}} - U\pts{\mathbf{x}^{3N}}
\end{equation}
is always the energy difference between the high-particle-number
configuration and the low-particle-number configuration.

There is an index permutation issue, though, about exactly what
(\ref{DetailedBalanceRequirement2}) means.  We know there are $N!$
copies of a particular differential volume hypercube $d{\bf x}^{3N}=
d{\bf x}_1d{\bf x}_2...d{\bf x}_N$ that are energetically degenerate,
permuting only the position of indistinguishable atoms. Similarly, we know that $(N+1)!$ copies of the
differential volume hypercube $d{\bf x}^{3N+3}= d{\bf x}_1d{\bf
  x}_2...d{\bf x}_{N+1}$.  Are we allowing Monte Carlo transitions
between any pair of them, i.e. a total of $N!(N+1)!$ transitions in
labelled-atom space, or are we only adding/deleting the last atom in
the labelled-atom space without random permutation afterwards,
i.e. in total only $(N+1)N!=(N+1)!$ Monte Carlo transition bridges? 
Since with the same computational cost, building more Monte Carlo
bridges facilitate approaching equilibrium, the first interpretation
($N!(N+1)!$ transitions) is preferable. Thus, if we take $\{N,
\mathbf{x}^{3N}\}$ to mean the ``index-free'' collection of all $N!$
degenerate copies, where every copy in the $\{N, \mathbf{x}^{3N}\}$
set automatically share the weight in the collective, then
(\ref{DetailedBalanceRequirement2}) is simplified to
\begin{equation}
  \label{DetailedBalanceRequirement3}
 \frac{{\rm transition}\pts{\{N,\mathbf{x}^{3N}\}\rightarrow
     \{N+1,\mathbf{x}^{3N+3}\} } d{\bf x}_{\rm question} }{{\rm
     transition}\pts{\{N+1,\mathbf{x}^{3N+3}\}\rightarrow
     \{N,\mathbf{x}^{3N}\} } } \;=\; \frac{e^{\beta\pts{\mu - \Delta
       U}}d{\bf x}_{\rm question} }{ \lambda_{\mathrm{th}}^{3}},
\end{equation}
where the $\{N, \mathbf{x}^{3N}\}$ notation is philosophically closer
to the quantum mechanical interpretation for identical particles.

Requirement (\ref{DetailedBalanceRequirement3}) is not that different
from the standard Monte Carlo for the canonical ensemble. Given one is
at $N,\mathbf{x}^{3N}$, one can attempt to insert (probability $a^+$)
/ accept insertion, or attempt to delete ($a^-=1-a^+$) / accept
deletion (\cite{OKeeffeRO07}). Both types of attempts would involve
computational cost, and rejection of either type of attempts would
mean wasted computations, and therefore $a^+/a^-$ may be chosen to
optimize performance, i.e. speed of approaching chemical equilibrium,
and efficiency in computing the thermodynamic averages of measurables.

Within the $a^+$ attempt branch, there is a question of where to
insert.  Again in the spirit of the simplest incarnation, we can
choose ``anywhere in the supercell, equally'', and therefore the
attempt probability to $d{\bf x}_{\rm question}$ is $(a^+/V) d{\bf
  x}_{\rm question}$, representing a spatially uniform prior. It does
not have to be this way.  If we have advanced screening information,
we could use umbrella sampling to tune this attempt probability
(indeed, with the domain decomposition scheme to come later, such
issue could arise). But right now let us choose the simplest insertion
prior.

Within the $a^-$ attempt branch, we can also choose ``any of the atoms
in the supercell, equally''. It does not have to be this way. If we
have advanced screening information, we could may umbrella sampling to
tune this attempt probability also.  But the $a^-/N$ prior (or
$a^-/(N+1)$ for the high-particle-number configuration) does lead to
the simplest proof of detailed balance.  Therefore, with these
simplest insertion/deletion priors, the requirement
(\ref{DetailedBalanceRequirement3}) is converted to
\begin{equation}
  \label{DetailedBalanceRequirement4}
  \frac{(a^+ / V)
    {\rm acceptance}\pts{\{N,\mathbf{x}^{3N}\} \rightarrow
   \{N+1,\mathbf{x}^{3N+3} \}}  }{(a^- / (N+1)) {\rm
     acceptance}\pts{\{N+1,\mathbf{x}^{3N+3}\}\rightarrow \{N,\mathbf{x}^{3N}\}
 } } \;=\;
 \frac{e^{\beta\pts{\mu -
     \Delta U}}}{
   \lambda_{\mathrm{th}}^{3}},
\end{equation}
where $d{\bf x}_{\rm question}$ is cancelled out, or
\begin{equation}
  \label{DetailedBalanceRequirement5}
  \frac{    {\rm acceptance}\pts{\{N,\mathbf{x}^{3N}\} \rightarrow
   \{N+1,\mathbf{x}^{3N+3} \}}  }{ {\rm
     acceptance}\pts{\{N+1,\mathbf{x}^{3N+3}\}\rightarrow \{N,\mathbf{x}^{3N}\}
 } } \;=\;
 \frac{a^- V e^{\beta\pts{\mu -
     \Delta U}}}{a^+ (N+1)
   \lambda_{\mathrm{th}}^{3}}.
\end{equation}
The Metropolis Monte Carlo dichotomy\cite{metropolis_equation_1953} is
then used to achieve (\ref{DetailedBalanceRequirement5}) literally, by
making one of the ${\rm acceptance}\pts{\{N,\mathbf{x}^{3N}\}
  \rightarrow \{N+1,\mathbf{x}^{3N+3} \}}$, ${\rm
  acceptance}\pts{\{N+1,\mathbf{x}^{3N+3}\}\rightarrow
  \{N,\mathbf{x}^{3N}\} }$ unity, and the other $\le 1$, depending on
the sign of the RHS.  So the standard GCMC algorithm is just
\begin{equation}
    \label{InsertionAcceptanceRate}
  {\rm acceptance}\pts{\{N,\mathbf{x}^{3N}\}
    \rightarrow \{N+1,\mathbf{x}^{3N+3} \}} \;=\;
  \left\{\begin{array}{cc}
1, &  \frac{a^- V e^{\beta\pts{\mu -
     \Delta U}}}{a^+ (N+1)
   \lambda_{\mathrm{th}}^{3}} \ge 1,\\
\frac{a^- V e^{\beta\pts{\mu -
     \Delta U}}}{a^+ (N+1)
   \lambda_{\mathrm{th}}^{3}}, & \frac{a^- V e^{\beta\pts{\mu -
     \Delta U}}}{a^+ (N+1)
   \lambda_{\mathrm{th}}^{3}} < 1,
\end{array}\right.
\end{equation}
and
\begin{equation}
  {\rm
     acceptance}\pts{\{N+1,\mathbf{x}^{3N+3}\}\rightarrow \{N,\mathbf{x}^{3N}\}
 }  \;=\;
  \left\{\begin{array}{cc}
1, &  \frac{a^+ (N+1)
   \lambda_{\mathrm{th}}^{3}}{a^- V e^{\beta\pts{\mu -
     \Delta U}}} \ge 1,\\
\frac{a^+ (N+1)
   \lambda_{\mathrm{th}}^{3}}{a^- V e^{\beta\pts{\mu -
     \Delta U}}}, & \frac{a^+ (N+1)
   \lambda_{\mathrm{th}}^{3}}{a^- V e^{\beta\pts{\mu -
     \Delta U}}} <1,
\end{array}\right.
\end{equation}
or equivalently,
\begin{equation}
  \label{DeletionAcceptanceRate}
  {\rm
     acceptance}\pts{\{N,\mathbf{x}^{3N}\}\rightarrow \{N-1,\mathbf{x}^{3N-3}\}
 }  \;=\;
  \left\{\begin{array}{cc}
1, &  \frac{a^+ N
   \lambda_{\mathrm{th}}^{3}}{a^- V e^{\beta\pts{\mu -
     \Delta U}}} \ge 1,\\
\frac{a^+ N
   \lambda_{\mathrm{th}}^{3}}{a^- V e^{\beta\pts{\mu -
     \Delta U}}}, & \frac{a^+ N
   \lambda_{\mathrm{th}}^{3}}{a^- V e^{\beta\pts{\mu -
     \Delta U}}} <1,
\end{array}\right.
\end{equation}
where $\Delta U$ is always the energy difference between the
high-particle-number configuration and the low-particle-number
configuration.  It is seen from above that $a^+/a^-$, $1/V$ and $1/N$
are just {\em choices}, reflecting the simplest prior about how to
make moves.  There is nothing set in stone about them. As long as we
use these choices {\em consistently}, detailed balance can be
established.

It is clear from the derivations above that the $0 < a^+ < 1$
insertion attempt rate and $a^- = 1-a^+$ deletion attempt rate can be
arbitrarily tuned, and also the $(a^+/V) d{\bf x}_{\rm question}$ and
$a^-/N$ priors can be substantially changed.  Indeed, our domain
decomposition scheme below changes these prefactors of Monte Carlo
sampling for each domain, as $N_{\rm domain}/V_{\rm domain}$ can vary
from domain to domain, and does not have to be equal to the average
$N/V$.

\subsubsection{Domain Decomposition}

Imagine that our supercell is spatially partitioned into $d=1..D$
separate domains. Analytically, one can rearrange
(\ref{GrandPartitionFunction}) into the following form
\begin{align}
  \label{DomainDecompositionGrandPartitionFunction}
\mathcal{Q}\pts{T,\mu,V}=\mathcal{I}_1\cdots \mathcal{I}_D
\exp\left[ -\beta U\pts{\mathbf{x}^{3N_1},\cdots,\mathbf{x}^{3N_d}}\right],
\end{align}
where
\begin{align}
  \mathcal{I}_d \equiv \sum_{N_d=0}^{\infty} \int_{V_d}
  \frac{e^{\beta\mu N_d}}{N_d ! \lambda_{\mathrm{th}}^{3N_d}}
  d\mathbf{x}^{3N_d},
\end{align}
is an integration and summation operator. The $1/N_d!$ prefactor comes
from the combinatoric $N!/N_1! N_2! ... N_D!$ copies of assigning
labeled but identical nuclides to different domains, that yields the
same integral for (\ref{GrandPartitionFunction}). Namely, in the way
(\ref{GrandPartitionFunction}) was written, all particles can traverse
and live in all domains, but now in the new form
(\ref{DomainDecompositionGrandPartitionFunction}), within each
$\mathcal{I}_d$ only ``citizens'' of that domain can live and
contribute to the integral. Therefore the probability density of a
microstate while ignoring the labeling of particles inside each domain
is:
\begin{align}
\label{eq:pbarpar}
\bar{p}\pts{\{N_1, \mathbf{x}^{3N_1}\}, \cdots, \{N_D,
  \mathbf{x}^{3N_D}\}} = e^{\beta\mathcal{G}} \pts{\prod_{d = 1}^{D}
  \frac{d\mathbf{x}^{3N_d}}{\lambda_{\mathrm{th}}^{3N_d}}e^{\beta\mu
    N_d}} \exp\left[ -\beta
  U\pts{\mathbf{x}_1,\cdots,\mathbf{x}_D}\right].
\end{align}

Now let us modify the standard GCMC algorithm by changing the
priors. Every time that an insertion or deletion attempt trial is to
be performed, one of the domains, say $d$ is randomly chosen with a
probability $P_d$:
\begin{equation}
 \sum_{N_d=0}^{\infty} P_d \;=\; 1,
\end{equation}
$P_d$ can be chosen to be proportional to its volume $V_d$, for
example. Other priors may be chosen, but one should be careful in not
letting $P_d$ depending on $N_d$, which can dynamically change, that
unless proven, may break the detailed-balance requirement. A
time-constant $\{P_d\}$ distribution should always be fine, if
computational efficiency is of no concern.

The domain-$d$ insertion/deletion attempt rate is therefore $P_da^+_d$
/ $P_da^-_d$, and we can straightforwardly show that
\begin{equation}
    \label{DomainInsertionAcceptanceRate}
  {\rm acceptance}\pts{\{N_d,\mathbf{x}^{3N_d}\}
    \rightarrow \{N_d+1,\mathbf{x}^{3N_d+3} \}} \;=\;
  \left\{\begin{array}{cc}
1, &  \frac{a^-_d V_d e^{\beta\pts{\mu -
     \Delta U}}}{a^+_d (N_d +1)
   \lambda_{\mathrm{th}}^{3}} \ge 1,\\
\frac{a^-_d V_d e^{\beta\pts{\mu -
     \Delta U}}}{a^+_d (N_d+1)
   \lambda_{\mathrm{th}}^{3}}, & \frac{a^-_d V_d e^{\beta\pts{\mu -
     \Delta U}}}{a^+_d (N_d+1)
   \lambda_{\mathrm{th}}^{3}} <1,
\end{array}\right.
\end{equation}
and
\begin{equation}
  \label{DomainDeletionAcceptanceRate}
  {\rm
     acceptance}\pts{\{N_d,\mathbf{x}^{3N_d}\}\rightarrow \{N_d-1,\mathbf{x}^{3N_d-3}\}
 }  \;=\;
  \left\{\begin{array}{cc}
1, &  \frac{a^+_d N_d
   \lambda_{\mathrm{th}}^{3}}{a^-_d V_d e^{\beta\pts{\mu -
     \Delta U}}} \ge 1,\\
\frac{a^+_d N
   \lambda_{\mathrm{th}}^{3}}{a^-_d V_d e^{\beta\pts{\mu -
     \Delta U}}}, & \frac{a^+_d N_d
   \lambda_{\mathrm{th}}^{3}}{a^-_d V_d e^{\beta\pts{\mu -
     \Delta U}}} <1,
\end{array}\right.
\end{equation}
where $\Delta U$ is always the energy difference between the
high-particle-number configuration and the low-particle-number
configuration, and may depend on nearby domains, would preserve
detailed balance. While
(\ref{DomainInsertionAcceptanceRate}),(\ref{DomainDeletionAcceptanceRate})
is factually a different algorithm from
(\ref{InsertionAcceptanceRate}),(\ref{DeletionAcceptanceRate}), their
derivation follows exactly the same logic flow of the previous
section, just with a {\em different set of screening priors} for
making the next move. That is, in
(\ref{DomainInsertionAcceptanceRate}),(\ref{DomainDeletionAcceptanceRate})
all ``citizens'' and all volume elements of the same ``country''
(domain) are treated the same before the ``testing'' (evaluation) of
the potential (``presumed innocent''), but they are treated
differently from country to country, whereas in
(\ref{InsertionAcceptanceRate}),(\ref{DeletionAcceptanceRate}) all
``citizens'' and volume elements of the ``world'' (supercell) are
treated equally before the testing of the potential.  Even though
(\ref{DomainInsertionAcceptanceRate}),(\ref{DomainDeletionAcceptanceRate})
is factually a different algorithm from
(\ref{InsertionAcceptanceRate}),(\ref{DeletionAcceptanceRate}), as
long as being run consistently, both can provide the equilibrium
ensemble distribution (\ref{pdf:micro}). So in this sense, the domain
partition of the supercell and the $\{P_d, a^+_d/a^-_d\}$ are just
``gauge'' choices.  These gauge choices, however, can be used to
enhance the computational efficiency, because it is also clear from
the logic flow of the proof that the ``global citizen'' approach of
(\ref{InsertionAcceptanceRate}),(\ref{DeletionAcceptanceRate}) has no
reason to posses optimal efficiency.  Thus $\{V_d, P_d, a^+_d/a^-_d\}$
can be optimized, and when $V_d$ is taken to be quite small, it is
clear that the $P_d$ spatial distribution would amount to a screening
prior in umbrella sampling.  While we do not explore this degree of
freedom fully here in this paper, taking the simplest uniform $V_d,
P_d$ approach, the connection between domain decomposition and
umbrella sampling is noted.

In addition to statistical sampling efficiency, there is the critical
issue of efficient interatomic potential evaluation and load
balancing, well-known in parallel computing for discrete agent-based
simulations with short-range interactions.  To this end, we break down
a large supercell to smaller individual domains, whose size is chosen
according to the radial cutoff distance in the interatomic potential
and possible range of dynamic strain in the supercell
\cite{Li05a,Li05b}. If the atoms in two separate domains cannot
possibly interact with one another, energy difference would only
depend on the affected domain and nothing prohibits us from performing
$D$ simultaneous moves. Although the aforementioned assumption is
almost never the case, it is possible to choose a pattern of domains
such that the changes of potential energy due to any perturbation in
the domains are independent of one another.

In the limit of truly isolated, non-interacting domains, it is clear
that (\ref{DomainDecompositionGrandPartitionFunction}) will become a
product of domain-specific grand partition functions. Thus,
(\ref{DomainInsertionAcceptanceRate}),
(\ref{DomainDeletionAcceptanceRate}) can be interpreted as running
chemical equilibration between the constant-$\mu$ reservoir with each
domain indepedently, using the classic GCMC algorithm.  For load
balancing purposes, the frequency of attempting to equilibrate these
different domains may not need to be the same.  In problems involving
multi-phase chemical equilibrium, for instance, one may have a vapor
phase with very different $N_d/V_d$ with a solid phase that it should
be in equilibrium with.  The freedom to pick $P_d, a^+_d/a^-_d$ is in
effect an umbrella-sampling scheme based on spatial location for
load-balancing and sampling efficiency.  For detailed balance, $P_d,
a^+_d/a^-_d$ should not depend on $N_d$ turn-by-turn, as $N_d$
sustains microscopic fluctuations.  But by our derivations above, one
should be able to re-adjust $P_d, a^+_d/a^-_d$, say after every 1000
acceptances in that domain.  One may also re-partition the supercell
and redefine the domains for computational efficiency and expediency
(say in a parallel computer, the number of allocated computing nodes
may be forced to change from time to time), as long as this is not
done too frequently.

\subsection{Pattern selection algorithm}

As mentioned in the previous section, it is possible to construct a
pattern that involves non-interacting domains. Such a pattern can be
established by considering that these domains have to be at least $n
\times r_{\mathrm{C}}$ apart in any direction, where $r_{\mathrm{C}}$
is the cutoff radius of the interatomic potential and $n$ is an
integer determined by the type of the forcefield (see
\cite{sadigh_scalable_2012}).  For example, for pair potentials,
$n=1$; for embedded atom method (EAM), $n=2$; and for modified EAM,
$n=3$. Here, we describe how to construct such a pattern given that
the domains are predetermined by the MD part of the GCMC+MD method.

\begin{figure} [h!]
\centering
  \scalebox{1.0}{\includegraphics[]{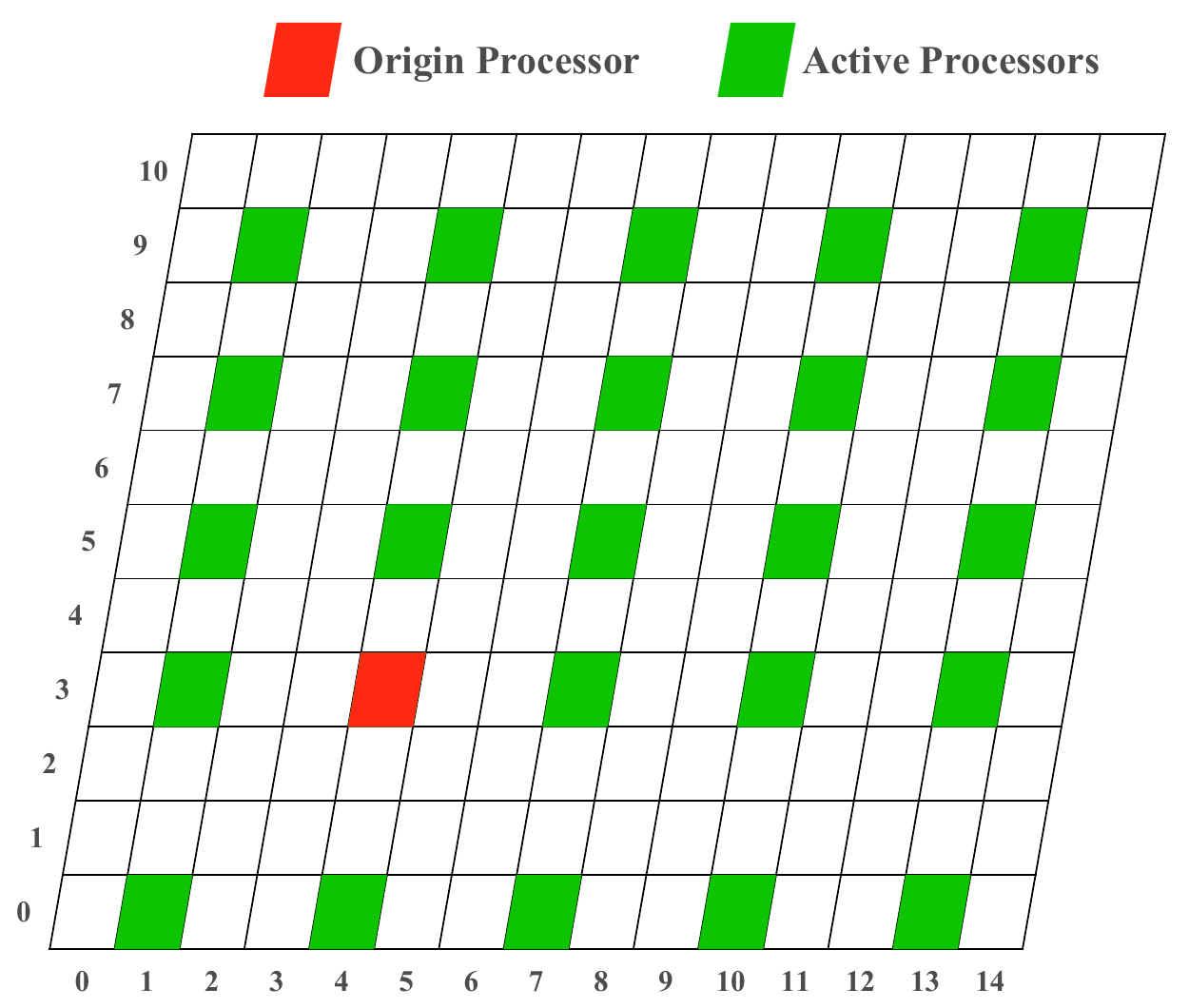}}
\caption{Schematic representation of subdomain selection pattern in two dimensions
for $\vec{n}=\{15,11\}$ and $\vec{m}=\{2,1\}$. Red and green subdomains denote origin 
and active processors, respectively.}
\label{fig:sd_pattrn}
\end{figure}

Suppose that our supercell is spanned by 
$n_0\times n_1\times n_2$ processors. Each processor
controls a domain determined by a domain decomposition
method. Fig. \ref{fig:sd_pattrn} depicts a simplified two dimensional schematic of such a supercell. Our first task is to 
determine how far apart our active domains should be.
In other words, we must determine how many domains
in each direction will at least cover $n\times r_{\mathrm{C}}$, namely 
integer vector $\vec{m}$. The best case scenario 
would be $\vec{m}=\{1,1,1\}$. In Fig. \ref{fig:sd_pattrn}, 
$\vec{m}=\{2,1\}$.
Let us define another integer
vector
\begin{align}
s_i=\lfloor n_i/\pts{m_i+1} \rfloor, \quad i=0,1,2,
\end{align}
where $\lfloor . \rfloor$ is the floor function. 
$\vec{s}$ defines the number of simultaneous trial moves 
in each direction at any given GCMC step. Therefore, the total 
number of simultaneous trial moves is $s_0\times s_1 \times s_2$.

At every step of GCMC, a processor is chosen at random.
This processor will serve as the ``origin,'' and its position
in the domain grid is denoted by an integer vector $\vec{o}$
(red domain in Fig. \ref{fig:sd_pattrn}). 
Now, all the processors that perform trial moves, namely 
``active,'' have to be determined (green domains in Fig. \ref{fig:sd_pattrn}).

Consider a processor whose position in the domain grid is denoted by the integer vector 
$\vec{p}$.  This processor is active if and only if 
\begin{align}
\sum_{i=0}^2\left[\pts{p_i+n_i-o_i} \mathrm{ mod }\, n_i\right] \mathrm{ mod } \pts{m_i+1}=0
\end{align}
where ``$\mathrm{mod}$'' denotes a modulo operation. The first modulo
operation takes into account the periodic boundary condition. The
second one ensures that active domains are non-interacting.  Once the
active processors are determined, they will perform trial moves
simultaneously.

\subsection{Potential energy difference calculation}

As was pointed out earlier, the most expensive part of GCMC or any
kind of atomistic simulation is the potential energy evaluation.  This
is due to the fact that potential energy and forces depend on the
interaction of atoms. However, in the case of short-range force
fields, the interactions of an atom with its surrounding can be
restricted to a volume within a cutoff radius. In MD, all the pairs
that are within a specified cutoff radius are included in a sparse
matrix, usually referred to as a neighborlist. The forcefield employs
the neighborlist to calculate energy/forces of the system. Since the
generation of the neighborlist is itself time-consuming, researchers
have come up with two major algorithms to speed things up: the
cell/linked list \cite{allen_computer_2017} and the Verlet algorithm
\cite{verlet_computer_1967}.

The cell/linked list algorithm is employed to limit the search volume
for finding all the pairs interacting with an atom. The Verlet
algorithm makes it possible to generate neighborlists less often.

In the case of GCMC, due to insertion or deletion moves, the
maintenance of a neighborlist has extra complication. Nonetheless, a
properly designed linked list will be relatively simple to update. In
our implementation of parallel GCMC, every processor has its own
linked list.  Whenever a trial move is to be performed, a neighborlist
for the affected atom will be instantly generated using the linked
list and passed to the forcefield to calculate the energy
difference. Next, the energy difference will determine whether the
move is accepted or rejected. If the move is accepted, the linked list
will be updated, accordingly.

In order to speed up even further, we modified the linked list
algorithm. In the traditional linked list algorithm, the supercell is
gridded by cubes (cells) with an edge of length $r_{\mathrm{C}}$. A
linked list would be generated to link all the atoms within a
cell. The basic premise of the algorithm is that when the neighborlist
of an atom inside one of these cells is to be generated, only the
atoms in the surrounding cells need to be searched, and these are
accessible via the said linked list. In three dimensions, the search
volume will be decreased from the volume of the whole supercell $V$ to
$27r_{\mathrm{C}}^3$.

\begin{figure} [h!]
\centering
  \scalebox{1.0}{\includegraphics[]{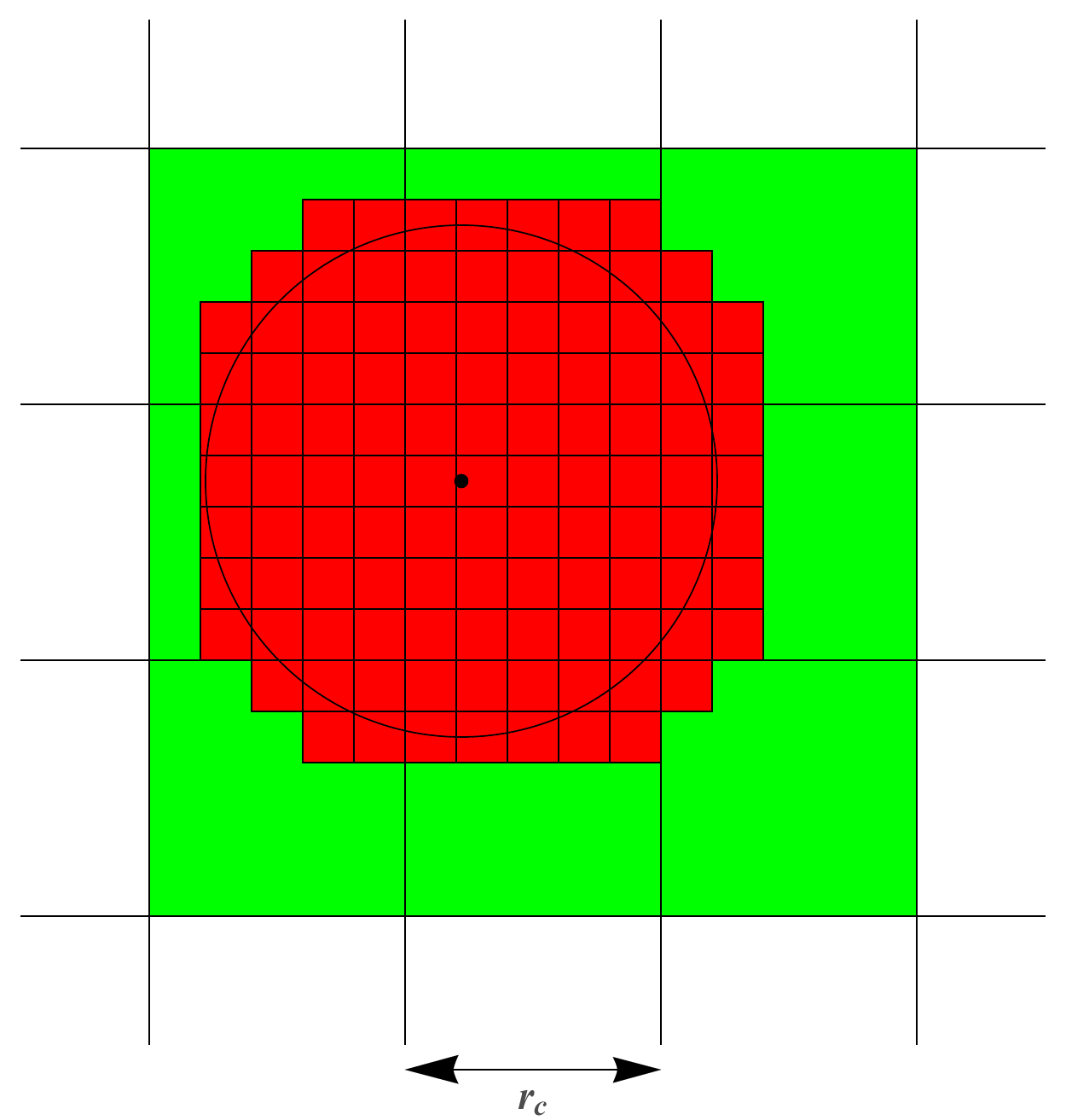}}
\caption{Classical linked list algorithm (green) versus proposed extension (red).} 
\label{fig:neigh_stencil}
\end{figure}

\begin{figure} [h!]
\centering
  \scalebox{1.5}{\includegraphics[]{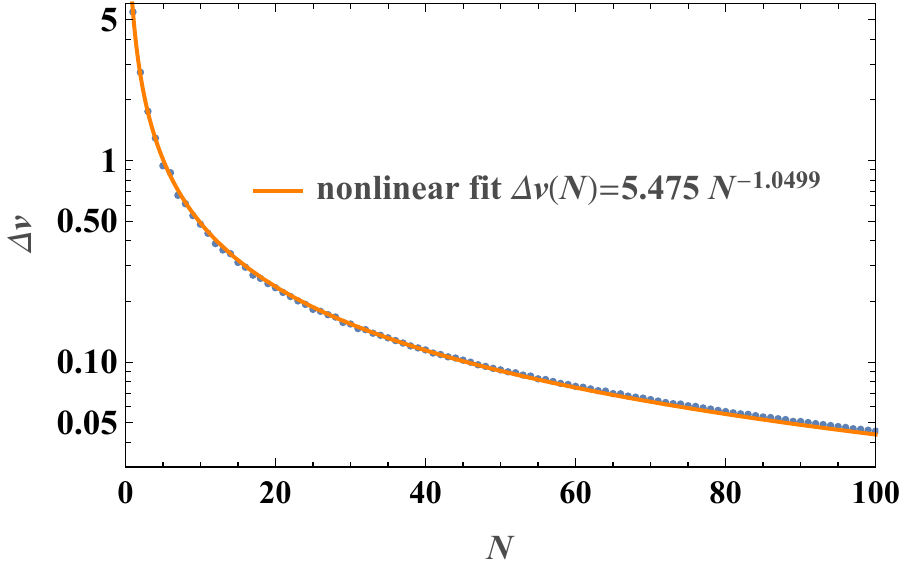}}
\caption{Relative excess search volume ($\Delta v$) with respect to 
number of discretization of cutoff radius ($N$).}
\label{fig:dv_vs_nbins}
\end{figure}

It is possible to reduce the search volume even further. In our
implementation the cell's edge length is $r_{\mathrm{C}}/m$, where $m$
is a positive integer to be set by the user, see
Fig. \ref{fig:neigh_stencil}. For two cells to interact, the minimum
possible distance between them must be smaller than or equal to
$r_{\mathrm{C}}$. Suppose cell $\Omega_i$ is labeled by a 3D vector
$\vec{s}_i$, which determines its position in the grid and the volume
it covers.  In other words, $\vec{x}\in \Omega_i$ if
\begin{align}
s_{i_\alpha}\frac{r_{\mathrm{C}}}{m}\leq x_\alpha < 
\pts{s_{i_\alpha}+1}\frac{r_{\mathrm{C}}}{m}, \quad \alpha=0,1,2.
\end{align}
One can easily show that for cells $\Omega_i$ and $\Omega_j$ 
to be interacting, the following must be true:
\begin{align}
\sum_{\alpha=0}^2\left[\min\pts{|s_{i_\alpha}
-s_{j_\alpha}|-1,0}\right]^2\leq m^2.
\end{align}
Using this relationship, a ``relative'' neighborlist for the cells is
created and will be utilized to build the neighborlist of a specific
atom in one of the cells. Fig. \ref{fig:dv_vs_nbins} shows relative
excess volume $\Delta v$ vs. the number of cutoff
discretizations. Here, relative excess volume is defined by the ratio
of search volume to the volume of the sphere with a radius
$r_{\mathrm{C}}$ minus one.  As the number of discretizations tends to
infinity, the excess volume will tend to zero. One might naively
conclude that the higher number of discretizations must inevitably
lead to better performance. However, this improving trend is true only
up to a point. The downside of increasing the number of
discretizations is an increase in the memory storage of the head array
of the linked list. If the storage becomes too large, it can lead to
cache pollution and indeed poor performance. Much like the Verlet
algorithm, in which the size of the shell must be chosen according to
the problem under study, the number of discretizations can be chosen
empirically by trial and error.

\section{Results}
\subsection{Scalability tests}

To demonstrate the performance and the scalability of our
implementation, several benchmark simulations were performed. All of
these tests were conducted on a Beowulf Linux cluster, where each of
the computational nodes contained two Intel Xeon Gold 6248 processors
with 20 cores. The code was compiled using a C++ GNU compiler and
\textbf{-O3} optimization flag.

These benchmarks simulated H absorption in a ferrite-phase Fe single
crystal. A sample consisting of $64 \times 64 \times 64$ Fe
body-centered-cubic unit cells was generated. To facilitate the
insertion of H, $0.1\%$ of atoms were randomly removed. The total
number of Fe (ferrite phase) atoms is approximately 524,000. The EAM
interatomic potential developed by Ramasubramaniam \textit{et al.}
\cite{ramasubramaniam_interatomic_2009} was used to model atomic
interactions.

The sample was in equilibrium with a reservoir that had a
$-2.4242~\mathrm{ eV}$ H chemical potential. The temperature was
$300~\kelvin$; in total, $10^6$ Monte Carlo insertion/deletion trials
were conducted.

\begin{figure} [h!]
\centering
  \scalebox{1.0}{\includegraphics[]{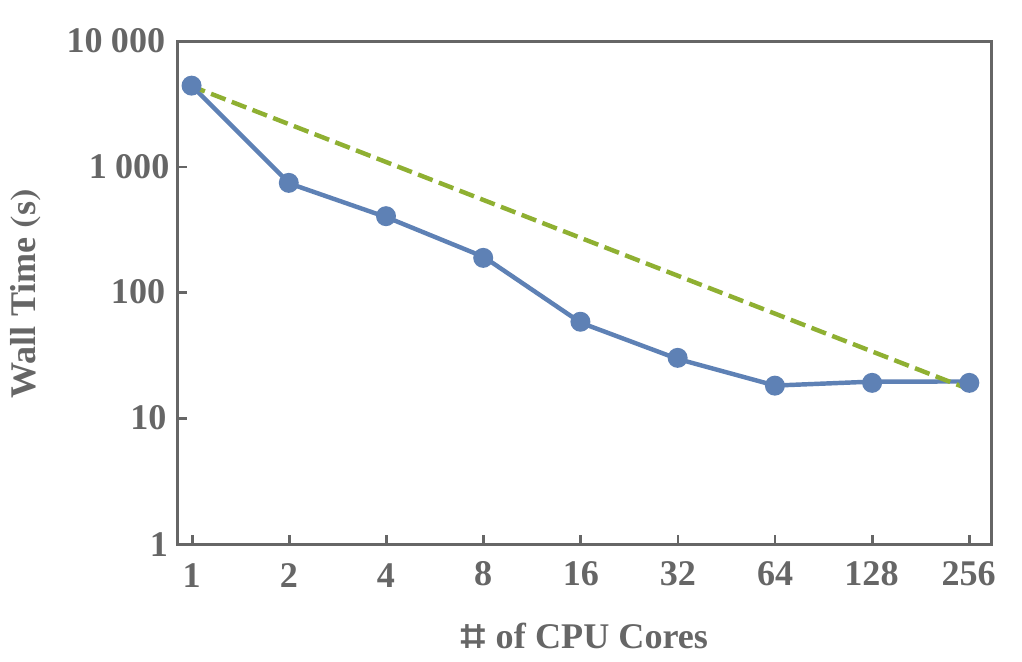}}
\caption{Total wall time as function of number of processors for $10^6$ MC trial moves of H in ferrite.} 
\label{fig:np_vs_wall}
\end{figure}

The tests were performed on 1, 2, 4, 8, 16, 32, 64, 128, and 256 CPU
cores. For each given number of processors, 4 tests were performed,
and the average wall time was recorded as the
result. Fig. \ref{fig:np_vs_wall} shows the total CPU wall-time with
respect to the number of processors, with the green dashed line being
the ideal scalability line. Overall, the trend looks like a typical
domain decomposition parallel application.

The biggest decrease in the computational time is the transition from
the serial execution (1 core) to the parallel execution on 2
cores. However, this decrease is not due to parallel MC moves. In
fact, the parallel MC moves will only take place starting from 16
processors. The main reason for the reduction in the computational
time below the 16 cores is the reduction of cache pollution. In other
words, due to the reduction of the number of atoms per processor,
fewer cache misses would occur in the energy calculation.

The second largest drop takes place on transitioning from 8 to 16
processors, i.e., the onset of parallel MC moves. The improvement in
computational time continues in a consistent manner up to 64
processors, reaching the start of a plateau. Increasing the number of
processors does not enhance the performance any further. In fact, it
is evident from the plot that the performance even slightly
suffers. This slight decrease in performance can be explained by the
overhead in inter-processor communication, which is the bottleneck of
performance beyond this point. In total, in comparison to current
widely used numerical libraries like LAMMPS \cite{plimpton_fast_1995},
for this problem, our framework is 250 times faster.

\subsection{Isothermal H absorption/desorption in a Ni polycrystalline sample with free surfaces}

As an example of application of the code, we chose isothermal H
absorption/desorption in a polycrystalline Ni sample with free
surfaces . This example was used in a previous work to study the
mobility of dislocations due to H charging \cite{koyama_origin_2020}.

\begin{figure} [h!]
\centering
  \scalebox{0.35}{\includegraphics[]{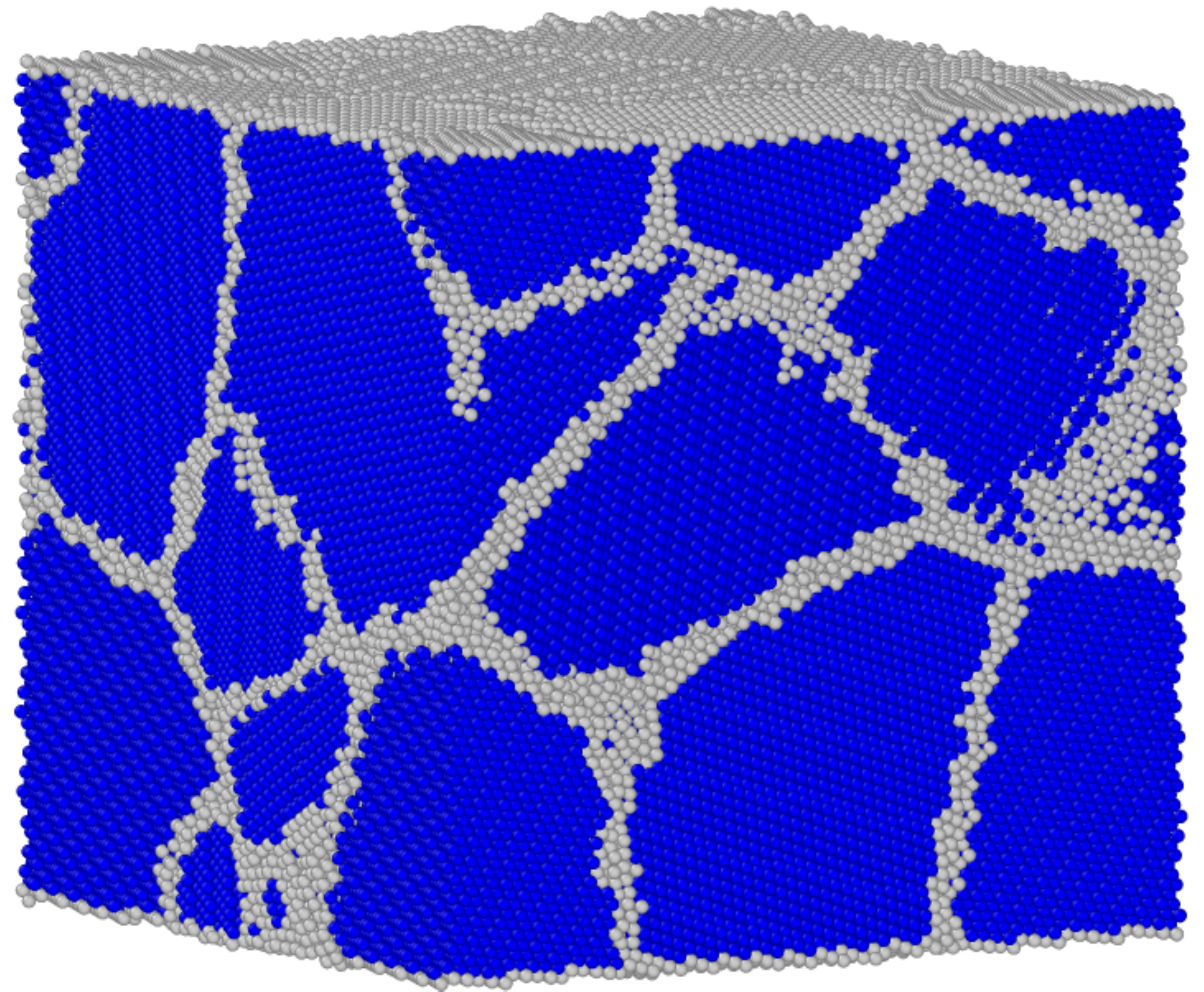}}
\caption{Polycrsytalline Ni sample with 8 randomly oriented grains for hybrid GCMC/MD simulations.}
\label{fig:micro}
\end{figure}

\begin{figure} [h!]
\centering
  \scalebox{1.0}{\includegraphics[]{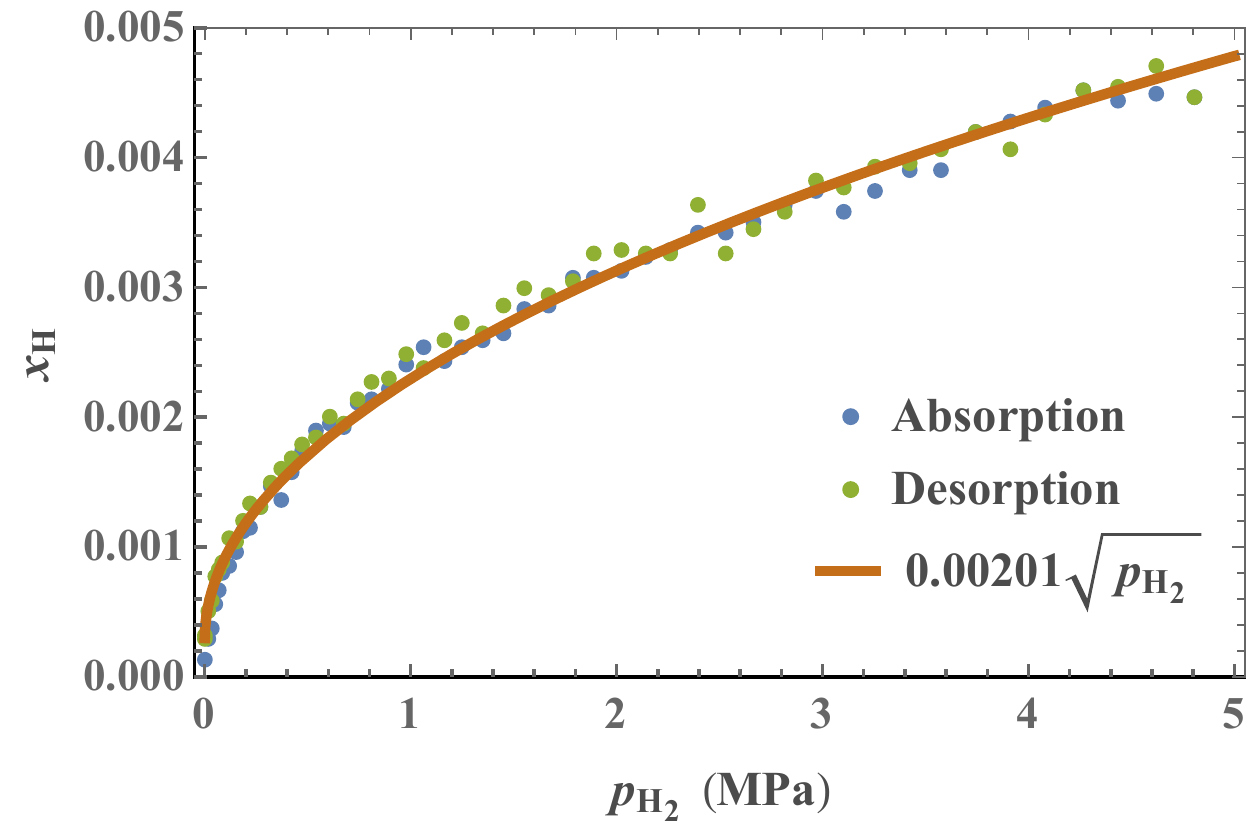}}
\caption{Isothermal curve for H absorption/desorption in the bulk of Ni polycrystalline sample. H concentration has a square root relationship with H pressure and follows Sieverts' law. H mainly locates in grain boundaries.}
\label{fig:bulk}
\end{figure}

\begin{figure} [h!]
\centering
  \scalebox{1.0}{\includegraphics[]{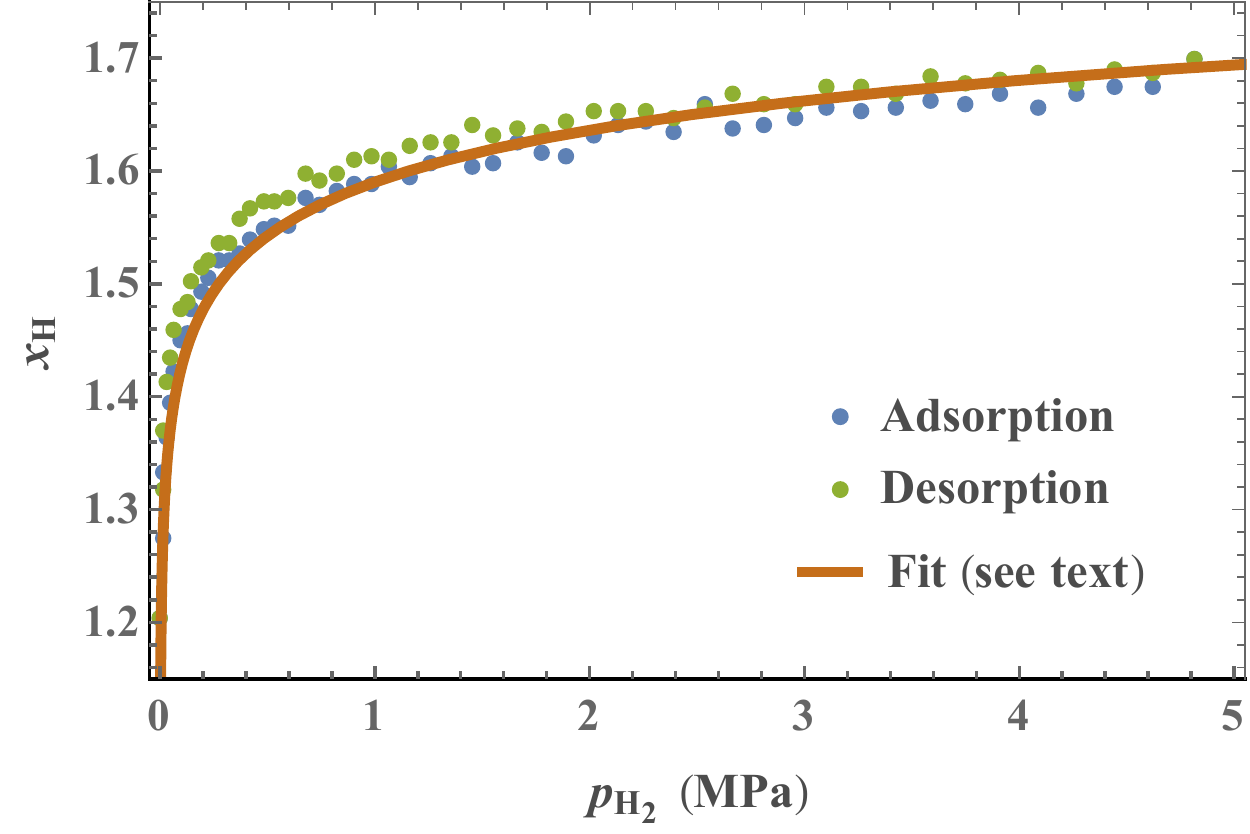}}
\caption{Isothermal curve for H absorption/desorption on the surface of Ni polycrystalline sample. H atom interactions influence the H concentration on the surface.}
\label{fig:surf}
\end{figure}

The sample has dimensions of 15 $\times$ 15 $\times$ 15
\nano\meter$^3$~and is periodic in the $x$ and $y$ directions. A 2.5
\nano\meter~vacuum layer was considered in the $z$ direction. The
sample contains about 300,000 atoms in eight randomly oriented grains,
which are approximately 7.5 \nano\meter~in diameter (see
Fig. \ref{fig:micro}).  For Ni and Ni-H interatomic interactions, the
EAM potential \cite{angelo_trapping_1995} was adapted.  Prior to
absorption, the Nos\'{e}-Hoover thermostat was used to maintain the
$NPT$ ensemble at room temperature $T$ = 300 \kelvin~and zero stress.
After stress relaxation, the sample was relaxed in a $\mu V T$
ensemble using hybrid GCMC+MD for $2~\nano\second$ with a time step of
$0.5~\femto\second$.  The chemical potential of H was set to
$\mu_0=-2.547~\mathrm{eV}$. After every 1000 steps of MD, 10,000 GCMC
trials were conducted. Thereafter, H chemical potential was varied
from $\mu_0$ to $\mu_1=-2.447~\mathrm{eV}$ in 50 increments of equal
pressure change $(P_1-P_0)/50$ from $P_0 \propto
\exp(2\mu_0/(k_{\mathrm{B}}T))$ to $P_1 \propto
\exp(2\mu_1/(k_{\mathrm{B}}T))$. At each increment, hybrid GCMC+MD was
conducted for $200~\pico\second$, using the scheme described
above. After the chemical potential reached $\mu_1$, the procedure was
reversed to return to the initial chemical potential.
Fig. \ref{fig:bulk} and Fig.  \ref{fig:surf} show the evolution of the
H-concentration in the bulk and the free surface of the
polycrystalline Ni. No hysteresis was observed in either case. In the
case of bulk atoms, H mainly locates in grain boundaries.  As
demonstrated, H concentration inside the bulk has a square root
relation with pressure and obeys Sieverts' law (see Supplementary
Discussion 1 for the derivation of the equation).

However, for the surface, the story is different. Due to the high
concentration of H on the surfaces, the ideal solution model with no
H-H interactions, i.e., the basis of Sieverts' law, is no longer
valid. However, employing the regular solution model for 2D, i.e. the
Fowler-Guggenheim adsorption isotherm model with lateral interaction
between H* and H*, where * means surface site, can capture the
behavior.  We can define species $\mathrm{A}$ and $\mathrm{B}$ of
regular solution as a surface site being occupied by H and a vacancy,
respectively.  The excess Gibbs free energy per site due to mixing is:

\begin{align}
\notag \Delta g\left(x\right)=&n_{\mathrm{s}}k_{\mathrm{B}}T\left[
\frac{x}{n_{\mathrm{s}}}\log\left(\frac{x}{n_{\mathrm{s}}}\right)
+\left(1-\frac{x}{n_{\mathrm{s}}}\right)\log\left(1-\frac{x}{n_{\mathrm{s}}}\right)
\right]\\
&+n_{\mathrm{s}}w \frac{x}{n_{\mathrm{s}}}\left(1-\frac{x}{n_{\mathrm{s}}}\right).
\end{align}
Here $x$ and $n_{\mathrm{s}}$ denote the concentration of H and 
number of H sites per Ni atom, respectively. 
$w$ is the effective interaction energy of the regular model. 
\begin{align}
w=w_{\mathrm{AB}}-\frac{w_{\mathrm{AA}}+w_{\mathrm{BB}}}{2}
\end{align}
Therefore,
\begin{align}
\mu_{\mathrm{H}}=\frac{\partial \Delta g}{\partial x}\Big|_{P,T}=k_{\mathrm{B}}T\log\left(\frac{x/n_{\mathrm{s}}}{1-x/n_{\mathrm{s}}}\right)+w\left(1-2\frac{x}{n_{\mathrm{s}}}\right),
\end{align}
As shown in details in Supplementary Discussion 1, since H is a diatomic gas, we can assume 
that its pressure is proportional to 
$\exp\left( 2\beta \mu_{\mathrm{H}} \right)$, leading to 
\begin{align}
P_{\mathrm{H_2}}\left(x\right)= C\left(\frac{x/n_{\mathrm{s}}}{1-x/n_{\mathrm{s}}}\right)^2e^{-4\beta wx/n_{\mathrm{s}}}
\end{align}
Based on the concentration curves, we conclude that
$n_{\mathrm{s}}=2$.  Fig. \ref{fig:surf} shows that the fit is in
excellent agreement with the values obtained from the
simulations. Based on our fitting, the effective interaction was
calculated to be almost zero for H in grain boundaries
(Langmuir–McLean isotherm), and $w=-0.22~\mathrm{eV}$ for H on free
surfaces.

\section{Discussion}
\color{black}

In conclusion, we present a hybrid GCMC/MD framework that can
efficiently simulate interstitial solid solution behavior in large
polycrystalline samples. We show that the parallelization of the code
is necessary for samples with large numbers of atoms .We provide two
applied case studies for H-absorption/desorption in a polycrystalline
Ni sample. Our analytical analysis was an excellent match with the
obtained numerical results.  The hidden parameters in the theory
(H-interaction energy) can now be extracted from our efficient
library. The framework has broad applications for simulation of
interstitial alloying elements such as C, H, and O in different
alloying systems and provides a new pathway to study the
diffusion-deformation mechanisms in these samples.


All data required to reproduce the findings during this study are
included in this Manuscript and Supplementary Information.  The code
is available for downloading at \texttt{https://github.com/sinamoeini}

\section{Acknowledgments}

The authors acknowledge the financial support from Timken (to
S.S.M.A.) and the Swiss National Science Foundation through Grant
P300P2$\_${171423} (to S.M.T.M.) and the U.S. Department of Energy
(DOE) Fuel Cell Technologies Office under award number DE-EE0008830
(to J.L.). Discussions with Doug Smith and R. Scott Hyde are
gratefully appreciated. The simulations reported were performed on a
local high-performance cluster. The authors also acknowledge the MIT
SuperCloud and Lincoln Laboratory Supercomputing Center for providing
(HPC and consultation) resources that have contributed to some of the
numerical results reported within this paper.

\section{Author Contributions}
S.S.M.A. and J.L. designed the project. S.S.M.A., J.L. and S.M.T.M. performed 
the analytical derivations. S.S.M.A. wrote the code.  S.S.M.A. and S.M.T.M.  tested
the code and ran the examples. S.S.M.A., S.M.T.M. and J.L. wrote the
paper.

\section{Competing Interests}
The authors declare no competing interests.

\bibliography{ref}
\bibliographystyle{naturemag}

\newpage

\newpage

\newpage

\newpage

\newpage

\newpage

\newpage
\section{Supplementary Information Document:}

\section{Supplementary Discussions 1| Analytical model of concentration-pressure relationships in Ni polycrystalline samples}

Developing an analytical model for the relationship between a concentration and a pressure of H in different defects requires several steps of analysis. We started by developing the free energy of $\mathrm{H_2}$ molecules with the help of H diatomic partition functions. The derivative of the H free energy with respect to the volume and number of H defines the $P-\mu$ relationship. Next, we developed the Gibbs free energy of the polycrystalline bulk Ni sample by incorporating the enthalpy and entropy contributions of Ni, vacancy, and H atoms. The H concentration-chemical potential $x-\mu$ relationship will be obtained from this equation. Substitution of $\mu$ from the thermodynamic relationship establishes $x-P$. This is in fact Sieverts' law which was used to fit the H concentration in grain boundaries. As discussed in the article, the mixing enthalpy term should also be considered for H on the surface of Ni polycrystalline samples; in this case H-H interactions due to the high concentration cannot be neglected. The following presents the details of our derivation.

\subsubsection{Partition function of H molecule ($\mathrm{H_2}$) }
The partition function of H molecules consists of the partition functions contributed by H-H bonding and translational, vibrational, and rotational modes as follows:

\begin{align}
z=z_{\mathrm{trans.}}z_{\mathrm{bond.}}z_{\mathrm{vib.}}z_{\mathrm{rot.}}
\end{align}

\subsubsection{Bonding}
The H-H bonding partition function is
\begin{align}
z_{\mathrm{bond.}}=e^{-\beta E_{\mathrm{b}}},
\end{align}
where $E_{\mathrm{b}}$ is the formation energy of the molecule at zero temperature, and $\beta$ is $(k_{\mathrm{B}}T)^{-1}$. Let us define a characteristic temperature as follows:
\begin{align}
\Theta_{\mathrm{b}}=-\frac{E_{\mathrm{b}}}{k_{\mathrm{B}}}
\end{align}
Therefore 
\begin{align}
z_{\mathrm{bond.}}=\exp\left(\frac{\Theta_{\mathrm{b}}}{T}\right),
\end{align}
for H molecules 
\begin{align}
E_{\mathrm{b}}=-4.73832 \mathrm{eV}\Rightarrow \Theta_{\mathrm{b}}=54985.9^\circ\mathrm{K}.
\end{align}

\subsubsection{Translational mode}
Similar to the ideal gas 
\begin{align}
z_{\mathrm{trans.}}=\frac{V}{\lambda_{\mathrm{th}}^3},
\end{align}
where $V$ denotes the volume of the system and 
\begin{align}
\lambda_{\mathrm{th}}=\frac{h}{\sqrt{2\pi 2 m_{\mathrm{H}}k_{\mathrm{B}}T}}.
\end{align}
Here, $h$, $k_{\mathrm{B}}$, $m_{\mathrm{H}}$, and $T$ are the Planck constant,  Boltzmann constant, H atomic mass, and temperature, respectively. Please note that this $\lambda_{\mathrm{th}}$ differs from the one for H atoms by a factor of $2$ behind $m_{\mathrm{H}}$.

\subsubsection{Vibrational mode}
The vibrational mode of diatomic gas must be dealt with from the perspective of a quantum harmonic oscillator.
\begin{align}
z_{\mathrm{vib.}}=\sum_{n=0}^{\infty}e^{-\beta\hbar\omega\left(n+1/2\right)}=\frac{1}{2\sinh\left( \beta \hbar \omega/2\right)},
\end{align} 
where
\begin{align}
\hbar=\frac{h}{2\pi}.
\end{align} 
To keep the notation consistent 
\begin{align}
z_{\mathrm{vib.}}=\frac{1}{2}\mathrm{cosech}\left( \frac{\beta h \omega}{4\pi}\right),
\end{align} 
or even better we can write it as follows: 
\begin{align}
z_{\mathrm{vib.}}=\frac{1}{2}\mathrm{cosech}\left( \frac{\Theta_{\mathrm{v}}}{T}\right),
\end{align} 
where 
\begin{align}
\Theta_{\mathrm{v}}=\frac{h\omega}{4\pi k_{\mathrm{B}}}.
\end{align} 

For H molecules 
\begin{align}
\omega \approx 8.80803 \times 10^{14} \mathrm{S^{-1}}\Rightarrow \Theta_{\mathrm{v}}=3363.89^\circ\mathrm{K}.
\end{align} 

\subsubsection{Rotational mode}
Like the vibrational mode, the rotational mode must also consider quantum mechanics.

\begin{align}
z_{\mathrm{rot.}}=\sum_{l=0}^{\infty}\left(2l+1\right)\exp\biggl[-\frac{\hbar^2 l\left(l+1\right)}{2 I k_{\mathrm{B}}T} \biggr]
\end{align} 
At high temperature limits, the series can be calculated using the integration as:
\begin{align}
z_{\mathrm{rot.}}\approx \frac{2Ik_{\mathrm{B}}T}{\hbar^2}.
\end{align} 
However, at low temperatures, only the first two terms of the series are considered. Thus, 
\begin{align}
z_{\mathrm{rot.}}\approx \exp\biggl[-\frac{\hbar^2}{2 I k_{\mathrm{B}}T} \biggr]+3\exp\biggl[-\frac{2\hbar^2}{2 I k_{\mathrm{B}}T} \biggr]+\cdots.
\end{align}
Similarly to the vibrational mode, we can define a characteristic temperature as
\begin{align}
\Theta_{\mathrm{r}}=\frac{h^2}{8\pi^2I k_{\mathrm{B}}}.
\end{align}
For H, 
\begin{align}
I=2.84373\times 10^{-29} \mathrm{eVS^2}\Rightarrow = \Theta_{\mathrm{r}}=88.3977 ^{\circ} \mathrm{K}.
\end{align}
Since $\Theta_{\mathrm{r}}$ is much lower than the room temperature, for our purpose, the high temperature limit is more appropriate.

Finally, we can write the partition function of H molecules as follows: 
\begin{align}
z=&V\bar{z}\left(T\right),
\end{align}
\begin{align}
\bar{z}_q\left(T\right)=&\frac{\sqrt{2}}{\lambda_{\mathrm{th}}^3}\exp\left(\frac{\Theta_{\mathrm{b}}}{T}\right)\frac{T}{\Theta_{\mathrm{r}}}\mathrm{cosech}\left( \frac{\Theta_{\mathrm{v}}}{T}\right),
\end{align}
\begin{align}
\bar{z}_c\left(T\right)=&\frac{\sqrt{2}}{\lambda_{\mathrm{th}}^3}\exp\left(\frac{\Theta_{\mathrm{b}}}{T}\right)\frac{T^2}{\Theta_{\mathrm{r}}\Theta_{\mathrm{v}}},
\end{align}

where
\begin{align}
\bar{z}\left(T\right)=&\frac{T^{3/2}}{\bar{\lambda_{\mathrm{th}}}^3}\exp\left(\frac{\Theta_{\mathrm{b}}}{T}\right)\frac{1}{2}\mathrm{cosech}\left( \frac{\Theta_{\mathrm{v}}}{T}\right)\frac{T}{\Theta_{\mathrm{r}}},
\end{align}
and 
\begin{align}
\notag\bar{\lambda_{\mathrm{th}}}=&12.2961\mathrm{\AA}\sqrt{^{\circ}\mathrm{K}},\\
\notag\Theta_{\mathrm{b}}=&54985.9^{\circ} \mathrm{K},\\
  \notag\Theta_{\mathrm{v}}=&3363.89^{\circ} \mathrm{K},
\end{align}
and 
\begin{align}
\notag\Theta_{\mathrm{r}}=&88.3977^{\circ} \mathrm{K}.
\end{align}

\subsubsection{Free energy of $\mathrm{H_2}$ molecules}
For $\mathrm{H_2}$ gas consisting of N  molecules, the free energy density is as follows:
\begin{align}
F&=-N k_{\mathrm{B}} T\log\left(\frac{V\bar{z}\left(T\right)}{N}\right)-N k_{\mathrm{B}} T,\\
p_{\mathrm{H_2}}&=-\frac{\partial F}{\partial V}\bigg|_{T,N}=nk_{\mathrm{B}}T,\\
S_{\mathrm{H_2}}&=-\frac{\partial F}{\partial T}\bigg|_{V,N}=Nk_{\mathrm{B}}\biggl[1-\log\left(n\right) +\frac{\partial }{\partial T} T\log\bar{z}\left(T\right) \biggr],\\
\mu_{\mathrm{H_2}}&=\frac{\partial F}{\partial N}\bigg|_{T,V}=-k_{\mathrm{B}} T\log\left(\frac{\bar{z}\left(T\right)}{n}\right),
\end{align}
where 
\begin{align}
n=\frac{N}{V}.
\end{align}
\subsubsection{Pressure-chemical potential ($P-\mu$) relationship}
Let us express everything in terms of pressure to extract the ($P-\mu$) relationship as:
\begin{align}
  \label{eq:mup}
\frac{\mu_{\mathrm{H}}}{k_{\mathrm{B}}T}&=-\frac{1}{2}\log\left(\frac{k_{\mathrm{B}} T\bar{z}\left(T\right)}{p_{\mathrm{H_2}}}\right).
\end{align}




\subsubsection{Free energy for a hydrogenated Ni polycrystalline sample}

Considering $n_{\mathrm{s}}$  as the number of possible H sites per Ni atom and $x$ as the H concentration $x = N_{\mathrm{H}}/N_{\mathrm{Ni}}$, the free energy of the hydrogenated sample is:

\begin{equation}
   \label{eq:1}
F = N f_{\mathrm{Ni}} + xN \Delta u_{\mathrm{H}} + n_{\mathrm{s}}Nk_{\mathrm{B}}T[\frac{x}{n_{\mathrm{s}}} \ln(\frac{x}{n_{\mathrm{s}}}) + (1-\frac{x}{n_{\mathrm{s}}})\ln(1-\frac{x}{n_{\mathrm{s}}})], 
\end{equation}

where $f_{\mathrm{Ni}}$ and $\Delta u_{\mathrm{H}}$ are Ni and H enthalpy, respectively. To calculate $\mu_{\mathrm{H}}$, a partial derivative with respect to the number of H atoms $xN$ should be performed:

\begin{equation}
\mu_{\mathrm{H}} = \frac{\partial F}{\partial (xN)} =  \frac{\partial F}{N\partial x}= \Delta u_{\mathrm{H}} + \frac{k_{\mathrm{B}}T n_{\mathrm{s}}N}{N}[ \frac{1}{n_{\mathrm{s}}}\ln(\frac{x}{n_{\mathrm{s}}}) + \frac{\frac{1}{n_{\mathrm{s}}}\frac{x}{n_{\mathrm{s}}}}{\frac{x}{n_{\mathrm{s}}}} - \frac{1}{n_{\mathrm{s}}} \ln(1-\frac{x}{n_{\mathrm{s}}})-\frac{1}{n_{\mathrm{s}}}]
\end{equation}
\[=  \Delta u_{\mathrm{H}} + k_{\mathrm{B}}T n_{\mathrm{s}}(\frac{1}{n_{\mathrm{s}}}(\ln(\frac{\frac{x}{n_{\mathrm{s}}}}{1-\frac{x}{n_{\mathrm{s}}}}))). \]
By substitution of $\mu$ with H pressure $P$ from Eq.~\ref{eq:mup}, we have:
\begin{equation}
\frac{\mu_{\mathrm{H}}}{k_{\mathrm{B}}T} = \ln\sqrt{\frac{P_{\mathrm{H_2}}}{k_{\mathrm{B}}T\bar{z}(T)}} = \frac{\Delta u_{\mathrm{H}}}{k_{\mathrm{B}}T} + \ln(\frac{\frac{x}{n_{\mathrm{s}}}}{1-\frac{x}{n_{\mathrm{s}}}})).
\end{equation}

\subsubsection{Sieverts' law and H in grain boundaries}
In the case of  $(x \ll 1)$, which is the situation for H in grain boundaries, the above equation becomes more simplified as
\begin{equation}
\ln\frac{\sqrt{\frac{P_{\mathrm{H_2}}}{k_{\mathrm{B}}T\bar{z}(T)}}}{\frac{x}{n_{\mathrm{s}}}} = \frac{\Delta u_{\mathrm{H}}}{k_{\mathrm{B}}T}. 
\end{equation}

This is in fact Sieverts' law, which shows that the H concentration $x$ is proportional to the square root of the pressure $P_{\mathrm{H_2}}$. This relationship is valid for the H in Ni grain boundaries.

\begin{equation}
x \propto A\sqrt{P_{\mathrm{H_2}}},
\end{equation}
where A is a constant value. Fitting Sieverts' law with our simulations, A is determined for the H in grain boundaries. 
\subsubsection{H on free surfaces}
Due to the high concentration of H atoms on the surfaces, the previous assumption of $x \ll 1$ is no longer valid. For this case, the H interactions should be also considered in the free energy. The interaction energy is as follows:

\begin{equation}
w = w_{\mathrm{AB}} - \frac{1}{2}(w_{\mathrm{AA}} + w_{\mathrm{BB}}),
\end{equation}
 where $\mathrm{A}$ denotes the interstitial sites filled with H, and $\mathrm{B}$ indicates vacant interstitial sites. Therefore, in this problem, only $w_{\mathrm{AA}}$ has a non-zero value. Extending Eq.~\ref{eq:1} by considering this energy, the free energy of the free surfaces containing H atoms is:  
 
\begin{equation}
F = N f_{\mathrm{Ni}} + xN \Delta u_{\mathrm{H}} + n_{\mathrm{s}}Nk_{\mathrm{B}}T[\frac{x}{n_{\mathrm{s}}} \ln(\frac{x}{n_{\mathrm{s}}}) + (1-\frac{x}{n_{\mathrm{s}}})\ln(1-\frac{x}{n_{\mathrm{s}}})]+ n_{\mathrm{s}}N\frac{x}{n_{\mathrm{s}}}(1-\frac{x}{n_{\mathrm{s}}}) w.
\end{equation}
 
The derivative of the free energy with respect to the total number of H atoms has an additional term compared to the one for grain boundaries as follows:

\begin{equation}
\mu_{\mathrm{H}} = \frac{\partial F}{\partial (xN)} =  \Delta u_{\mathrm{H}} + k_{\mathrm{B}}T n_{\mathrm{s}}(\frac{1}{n_{\mathrm{s}}}(\ln(\frac{\frac{x}{n_{\mathrm{s}}}}{1-\frac{x}{n_{\mathrm{s}}}}))) + (1-\frac{2x}{n_{\mathrm{s}}})w.
\end{equation}

By substitution of $\mu_{\mathrm{H}}$ by the H pressure and simplification of this equation, we have:
\begin{equation}
CP_{\mathrm{H_2}}  =  (\frac{\frac{x}{n_{\mathrm{s}}}}{1-\frac{x}{n_{\mathrm{s}}}})^2 e^{2( \frac{\Delta u_{\mathrm{H}}}{k_{\mathrm{B}}T} + (1-\frac{2x}{n_{\mathrm{s}}})\frac{w}{k_{\mathrm{B}}T})},  
\end{equation}

\begin{equation}
 P_{\mathrm{H_2}}\left(x\right)= C\left(\frac{x/n_{\mathrm{s}}}{1-x/n_{\mathrm{s}}}\right)^2e^{-4\beta wx/n_{\mathrm{s}}},
\end{equation}

where C is a constant. In the case of $w = 0$, this equation transforms into Sieverts' law.

\end{document}